NOTE

# A note on the 3D Brans-Dicke theory

**T. G. do Prado, V. S. Mendes and M. C. Verges**

COENQ, UTFPR, Ponta Grossa, PR, Brazil
DEMAT, UEPG, Ponta Grossa, PR, Brazil
COINF, UTFPR, Ponta Grossa, PR, Brazil

E-mail: `thiagoprado@utfpr.edu.br`

**Abstract.** Since the evidence for an accelerated universe and the gap of 70 % in the total energy, collected by WMAP, search for alternatives for the general relativity is an important issue, for this theory is not suited for these new phenomena. A particular alternative is the Brans-Dicke theory which has being allowing inspiring results, for example, concerning k-essence type fields in 4 dimensions. However, this theory is almost unexplored in the context of the dimensional reduction of the theory in 3 dimensions. In this work, we address some problems in this dimensional reduction, namely, evaluation of the deceleration parameter of the universe described by the 3 dimensional Brans-Dicke with and without matter. In both cases, we see that it is not possible to consider the theory as a model of k-essence descrybing the dark energy, but it can be considered as descrybing the dark matter.

## 1. Introduction

The evidence for an accelerated expansion of the universe as observed by WMAP and BOOMERANG [1, 2] lighted the discussion of eventual limits of the general relativity and the options for alternative theories of gravity, and among them we can cite scalar-tensor theories like supergravity, Kaluza-Klein theories, dual string theories, etc [3, 4]. This explains the recent interest in a particular scalar-tensor theory, namely Brans-Dicke theory [5, 6], proposed in the early sixties in order to incorporate the principle of Mach and the hypthesis of Dirac [7], namely an eventual variation with time of the Newton's gravitational constant $G$.

Most of the works which have been published up to now consider 4 flat dimensions, and some of them have tried to associate the scalar field of the Brans-Dicke theory as quintessence [8], and as a type of k-essence [9]. Others have tried to find a solution for the observed accelerated expansion by means of a 5D Brans-Dicke theory without matter [10, 11].

Concerning 3 dimensions, a large study has been done in gravitational theories since the publication of [12], motivated by the fact that 3D theories avoid some complications found in higher dimensional theories [13, 14]. However, there is not a deep study of 3D scalar-tensor theories, and it would be interesting to find some results in this matter, and in particular in 3D Brans-Dicke theories. For instance, we can address some problems like the association of the scalar field of the Brans-Dicke theory to k-essence fields which models the dark energy, as done in [9]. This is one of the two problems we concern in this paper. Specifically, our objective in this paper is to find an expression for the expansion rate of the universe as described by the 3D Brans-Dicke theory, with mattter and without matter.

## 2. The 3D Brans-Dicke Theory

The Brans-Dicke theory of gravity in 3D will be described by the action

$$S = \frac{1}{16\pi} \int d^3x \sqrt{|g|} \left[ \phi R - \frac{\omega}{\phi} g_{ab} \nabla^a \phi \nabla^b \phi \right] + \int d^3x \sqrt{|g|} \; L_M, \qquad (1)$$

where $R$ is the curvature scalar associated with the 3D metric $g_{ab}$, $\phi$ is the scalar field of the Brans-Dicke theory, $\omega$ is a dimensionless parameter of the theory, and $L_M$ represents the lagrangian of the matter fields which does not depend on $\phi$ (this is necessary in order to preserve the weak equivalence principle [4]). Note that, in 3D, the field $\phi$ has dimensions of inverse of lenght.

The equations for the gravitational field $g_{ab}$ derived from (1) read

$$\begin{aligned} G_{ab} = \frac{8\pi}{\phi} T_{ab}^M + \frac{\omega}{\phi^2} &\left[ (\nabla_a \phi)(\nabla_b \phi) - \frac{1}{2} g_{ab} (\nabla_c \phi)(\nabla^c \phi) \right] \\ &+ \frac{1}{\phi} \left( \nabla_a \nabla_b \phi - g_{ab} \nabla^2 \phi \right), \end{aligned} \qquad (2)$$

where $\nabla^2 = \nabla^a \nabla_a$, and $T^M_{ab}$ is the energy momentum tensor associated with the matter fields. From the equation (1) we find that the field equation for the scalar field $\phi$ is given by

$$\nabla^c \nabla_c \phi = \frac{8\pi}{2\omega + 3} g^{ab} T^M_{ab}. \tag{3}$$

In this context, we identity the second term in the equation (2) as the energy momentum tensor associated with the BD scalar field $\phi$, and write

$$8\pi T^{BD}_{ab} = \frac{\omega}{\phi^2} \left[ (\nabla_a \phi)(\nabla_b \phi) - \frac{1}{2} g_{ab} (\nabla_c \phi)(\nabla^c \phi) \right] + \frac{1}{\phi} \left( \nabla_a \nabla_b \phi - g_{ab} \nabla^2 \phi \right). \tag{4}$$

In what follows, we consider the perfect fluid approximation for both the matter fields and the BD scalar field $\phi$. Then the energy momentum tensor associated with matter will be given by

$$T^M_{ab} = g_{ab} p_M + (\rho_M + p_M) u_a u_b, \tag{5}$$

where $p_M$ and $\rho_M$ represent the pressure and the energy density associated with the matter-fluid and $u_a$ are the velocities. Comparing this relation with equation (3) we see that the BD scalar field and this matter-fluid are related by

$$\nabla^c \nabla_c \phi = \frac{8\pi}{2\omega + 3} (3p_M - \rho_M). \tag{6}$$

As for the BD scalar field $\phi$, its non-minimal coupling with gravity does not allow us to write an expression similar to the equation (5), relating the associated pressure $p_{BD}$ and energy density $\rho_{BD}$ for the field $\phi$, these quantities being identified as the spatial and temporal components of the energy momentum tensor associated with the BD scalar field, given by the equation (4). However, they are related by the equation of state $p_{BD} = \omega_{BD} \rho_{BD}$, where $\omega_{BD}$ should not be confused with the paramenter $\omega$ of the Brans-Dicke theory.

Moreover, in what follows, we consider a particular BD scalar field which depends only with the cosmic time $t$, such that $\phi = \phi(t)$.

## 3. 3D Brans-Dicke cosmology in the absence of matter

Let the 3D spacetime be described by the FRLW metric given by

$$ds^2 = g_{ab} dx^a dx^b = dt^2 - a^2(t) \left[ \frac{dr^2}{1 - kr^2} + r^2 d\varphi^2 \right], \tag{7}$$

where $a(t)$ is a dimensionless scale factor, and $k$ is a parameter which can assume only one of three values: $-1$ (open universe), $0$ (flat universe) and $+1$ (closed universe). In this work we consider only the flat case, following the observations done by WMAP.

Then, observing the equation (4) we find that the pressure $p_{BD}$ and the energy density $\rho_{BD}$ associated with the BD scalar field are given by

$$\rho_{BD} = \frac{1}{8\pi}\left[\frac{\omega}{2}\left(\frac{\dot{\phi}}{\phi}\right)^2 - 2\frac{\dot{a}}{a}\frac{\dot{\phi}}{\phi}\right], \tag{8}$$

$$p_{BD} = \frac{1}{8\pi}\left[\frac{\omega}{2}\left(\frac{\dot{\phi}}{\phi}\right)^2 + \frac{\dot{a}}{a}\frac{\dot{\phi}}{\phi} + \frac{\ddot{\phi}}{\phi}\right]. \tag{9}$$

Now, if the BD scalar field is assumed as a perfect fluid, then the equations (8) and (9) are supplemented by the equation of state

$$\dot{\rho}_{BD} + 2\frac{\dot{a}}{a}(\rho_{BD} + p_{BD}) = 0. \tag{10}$$

Now, if we consider the flat metric from (7) in the equations (2) and (6) we find

$$\left(\frac{\dot{a}}{a}\right)^2 = 8\pi\rho_{BD}, \qquad \ddot{\phi} + 2\frac{\dot{a}}{a}\dot{\phi} = 0. \tag{11}$$

In order to solve this system of coupled equations, we consider the ansatz

$$\phi(t) = \frac{1}{G_0}a^n(t), \qquad a(t) = a_0\left(1 + \frac{\beta}{\sqrt{G_0}}t\right)^\alpha, \tag{12}$$

where $G_0$ denotes the present value for the Newton's gravitational constant and $\beta$ is a dimensionless parameter. If we substitute this ansatz in equations (11) we find

$$n = \frac{2 \pm \sqrt{4 + 2\omega}}{\omega}, \qquad \alpha = \frac{1}{n+2}. \tag{13}$$

Now, substitution of the ansatz in the equation of state (10) results

$$n = -1, \qquad \omega = -2,$$

which is in agreement with (13). As a consequence, we see that $\alpha = 1$ and

$$a(t) = a_0\left(1 + \frac{\beta}{\sqrt{G_0}}t\right),$$

Thus, we find that the isotropic and homogenous 3D universe where the only field present is the BD scalar field $\phi(t)$ undergoes a linear evolution with time. Besides, the BD scalar field decreases with time, as we can see from equation (12). Substituting these solutions in the equations (8) and (9) we find that $\rho_{BD} = \frac{1}{8\pi}\frac{a_0^2}{a^2}$, which means that the energy density associated with the BD scalar field decreases with time, as expected from the solution for $\phi$, and $p_{BD} = 0$. Thus, we conclude that $\omega_{BD} = 0$ and, considering the strong energy condition (see [15] for details), we find that the deceleration parameter as given by

$$q = \frac{2\pi}{H^2}\rho_{BD} = \frac{a_0^2}{4H^2a^2},$$

with $H = \dot{a}/a$ being the Hubble parameter, is positive. This results shows that the BD scalar field cannot be considered as k-essence or quintessence field describing the dark energy. For this to be the case, the pressure must be negative, which lead to a negative value for the parameter $\omega_{BD}$. On the other hand, the BD scalar field can be used as a k-essence field modeling the dark matter, since the energy density $\rho_{BD}$ has a behaviour analogous to the energy density $\rho_M$ associated with matter for a 3D Brans-Dicke theory.

## 4. 3D Brans-Dicke cosmology in the presence of matter

In this section we adress the problem of the cosmology derived from the 3D Brans-Dicke theory in the presence of matter in a flat background. As we said in section 2, the matter will be modeled as a perfect fluid with associated energy density $\rho_M$ and pressure $p_M$. Substituting the FRLW metric (7) in the field equations (2) and (3), we see that the resulting Friedmann equations, in this case, are given by

$$\left(\frac{\dot{a}}{a}\right)^2 = \frac{8\pi}{\phi}\rho_M + 8\pi\rho_{BD}, \tag{14}$$

$$\ddot{\phi} + 2\left(\frac{\dot{a}}{a}\right)\phi = \frac{8\pi}{(3+2\omega)}\rho_M. \tag{15}$$

If we require the conservation of the energy momentum tensor, then we see that the equations of state for the matter fields and for the BD scalar field $\phi$ are given by

$$\dot{\rho}_M + 2\left(\frac{\dot{a}}{a}\right)\rho_M = 0, \tag{16}$$

$$\dot{\rho}_{BD} + 2\left(\frac{\dot{a}}{a}\right)(\rho_{BD} + p_{BD}) = \frac{\dot{\phi}}{\phi^2}\rho_M. \tag{17}$$

The non-minimal coupling between the BD scalar field $\phi$ and gravity in the action (1) amounts to a mixed term in the equation of state for $\phi$ involving the matter density $\rho_M$, which makes its solution a non-trivial one. Also, we see that, since the BD scalar field $\phi$ is massless and does not interact with matter, the equation (16) preserves the weak equivalence principle. And, by integrating this equation (16), we obtain that the energy density $\rho_M$ is related to the scale factor $a(t)$ by means of $\rho_M = a_0^2/a^2$.

As in the previous section, now we propose the solutions for the equations (14) and (15) the ansatz

$$\phi(t) = \frac{1}{G_0}(1+\chi t)^\gamma, \qquad a(t) = a_0(1+\chi t)^\alpha, \tag{18}$$

where $\chi$ is some constant. Replacement of this ansatz in the coupled field equations (14) and (15) gives us the results

$$\alpha = \frac{3+2\omega}{4+4\omega}, \qquad \gamma = \pm\alpha\frac{\sqrt{12+10\omega}}{6+5\omega}. \tag{19}$$

Unlike in the scalar-vaccum configuration, now it is not possible to determine the parameter $\omega$ since the state equation (17) is coupled. If we replace the ansatz (18) in the general equations (8) and (9) we find the relations for the energy density and pressure associated with the BD scalar field are given by

$$\rho_{BD} = \frac{1}{8\pi}\gamma\left(\frac{\omega\gamma}{2} - 2\alpha\right)t^{-2}, \tag{20}$$

$$p_{BD} = \frac{1}{8\pi}\left[\alpha\gamma + \frac{\omega\gamma^2}{2} + \gamma(\gamma-1)\right]t^{-2}. \tag{21}$$

If the strong energy condition is considered, we find that the deceleration parameter of the 3D Brans-Dicke theory with a distribution of matter is given by

$$q = \frac{2\pi}{H^2}\left[\frac{\rho_M}{\phi} + (1+2\omega_{BD})\rho_{BD}\right] \quad (22)$$

with $\omega_{BD}$ being the ratio between the pressure and energy density associated with BD scalar field, and $H$ being the Hubble parameter. We observe that the first term in the right side is completely related with matter, and the second term is completely related with the BD scalar field.

## 5. Conclusions

Analysis of the behaviour of the parameter $\omega_{BD}$ as a function of the Brans-Dicke parameter $\omega$ shows that $\omega_{BD} > 0$ for all $\omega > -0.628$. Also, the asymptotics $\omega \to \infty$ occurs for $\omega_{BD} = 1$ which lead to the conclusion that, for large values of $\omega$, the 3D Brans-Dicke theory with matter is not a type of k-essence modeling dark energy, since the deceleration parameter $q$ is larger than zero. These results are interesting, since Hongsu Kim in [9] shows that the 4D Brans-Dicke theory actually is a type of k-essence theory.

To finalize, we can compare the 3D General Relativy with matter and the 3D Brans-Dicke theory without matter, and see that both lead to a universe which undergoes a linear evolution. This motivates us to employ the value 1 for $\alpha$ in the ansatz (18). In this case, we find that $\omega = -0.5$, and $\gamma = \pm 0.756$. The energy density associated with BD scalar field is found to be

$$\rho_{BD} = \frac{1}{8\pi}\gamma\left(\frac{\omega\gamma}{2} - 2\right)t^{-2}. \quad (23)$$

and the requirement of positive definiteness of this quantity [15] amounts to $\gamma = -0.756$ to be the only acceptable value for $\gamma$. Thus, the BD scalar field is decreasing with time. The pressure associated with BD scalar field is found to be

$$p_{BD} = \frac{1}{8\pi}\gamma^2\left(\frac{\omega}{2} + 1\right)t^{-2}. \quad (24)$$

Replacing $\gamma = -0.756$ and $\omega = -0.5$, in equations (23) and (24) we find $\rho_{BD} \approx 5.4475 \times 10^{-2}t^{-2}$ and $p_{BD} \approx 1.7056 \times 10^{-2}t^{-2}$. This implies $\omega_{BD} \approx 0,313$. We see that, for the Brans-Dicke model for a 3D universe with expands linearly with cosmic time, the deceleration parameter $q$ is positive. Thus, the universe is decelerating, which is in agreement with the result found for the behaviour of the universe without matter.

The results found in this paper consider only a flat background, a choice done in a observational basis. But the models with non-null curvature should not be discarded a priori. A further problem which is worth to consider is to measure the effect of the curvature in the deceleration parameter, and to compare with similar calculations for higher dimensional theories.